\documentclass[a4paper,
               ]{jacow}
\usepackage{pdfpages,multirow,ragged2e,doi} %
\makeatletter%
	\ifboolexpr{bool{xetex}}
	 {\renewcommand{\Gin@extensions}{.pdf,%
	                    .png,.jpg,.bmp,.pict,.tif,.psd,.mac,.sga,.tga,.gif,%
	                    .eps,.ps,%
	                    }}{}
\makeatother

\ifboolexpr{bool{xetex} or bool{luatex}} 
 {}                                      
 {\usepackage[utf8]{inputenc}}           

\usepackage[USenglish]{babel}

\ifboolexpr{bool{jacowbiblatex}}%
 {%
  \addbibresource{jacow-test.bib}
  \addbibresource{biblatex-examples.bib}
 }{}
\begin{document}

\title{New interferometric aperture masking technique for full transverse beam characterization using synchrotron radiation}

\author{U. Iriso\thanks{uiriso@cells.es}, L.~Torino, ALBA-CELLS Synchrotron, Cerdanyola del Valles, Spain \\
C.~Carilli, National Radio Astronomy Observatory (NRAO), Socorro - NM, USA \\
B.~Nikolic, University of Cambridge, Cambridge, UK  \\
N. Thyagarajan, CSIRO, Bentley WA, Australia  }
	 
	
\maketitle

\begin{abstract}
Emittance measurements using synchrotron radiation are usually performed using x-rays to avoid diffraction limits. Interferometric techniques using visible light are also used to measure either the horizontal or the vertical beam projection. Several measurements rotating the interferometry axis are needed to obtain a full beam reconstruction. 
In this report we present a new interferometric multi-aperture masking technique and data analysis, inspired by astronomical methods, that are able to provide a full 2-D transverse beam reconstruction in a single acquisition. Results of beam characterization obtained at ALBA synchrotron light source will also been shown.
\end{abstract}

\section{Introduction}

Two-dimensional (2D) transverse characterization of the particle beam is of fundamental importance in accelerators, as it provides a measurement of the beam emittance, one of the key figures of merit of an accelerators. 
The emittance describes the transverse distribution of particles, which is well described by a 2D Gaussian ellipse,  parameterised by its major axis, minor axis, and tilt angle. 
In synchrotron light sources, beam size measurements are based on the Synchrotron Radiation (SR) emitted by the electron beam. The most commonly used methods to measure the beam size are the x-ray pinhole and the double-aperture interferometry~\cite{Kube:Dipac07}.

The double-aperture interferometry was originally developed using visible light in~\cite{Mitsuhashi:1998em}, but it only provides a one dimensional measurement in the direction given by the baseline joining the aperture centers. Full 2D reconstruction can be obtained by rotating the aperture mask at different orientations~\cite{PhysRevAccelBeams.19.122801}, but at least 4 orientations are needed, meaning at least 4 inteferograms. A four-aperture square mask has been used to obtain the 2D source size with one single interferogram~\cite{Masaki:wr2006}, but such mask suffers from decoherence due to redundant sampling in Fourier space and it is also affected by the non-uniform illumination of the SR~\cite{Novokshonov:2017qsa}. 

These detrimental effects can be mitigated with a combination of
techniques widely used in astronomy~\cite{1987Natur.328..694H, 1980SPIE..231...18S}.  During the last year, we developed full 2D
characterization of the ALBA electron beam using a single visible light interferograms with multi-aperture masks (from 3 to
7-holes)~\cite{Carilli24}. In these masks, the vector baseline
between every pair of holes is unique, whence the term
\textit{Non-Redundant Aperture} (NRA).  This report describes the
technique, shows the results obtained by changing the beam coupling in the ALBA storage ring, and compares them with measurements using
well-established techniques at ALBA: the rotating
2-aperture mask (which provides beamsize measurements at the same location), or calculations using the emittances obtained with LOCO~\cite{SAFRANEK199727} and the x-ray pinhole~\cite{UI:IBIC22}.

\section{Experimental Setup}

\begin{figure}[!htb]
   \centering
   \includegraphics*[width=.95\columnwidth]{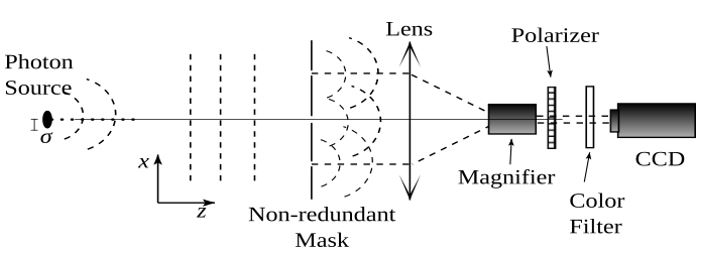}   
   \caption{Experimental set-up. The mask is exchanged to perform either NRA or rotating double-aperture inteferometry. }
   \label{fig:expsetup}
\end{figure}

Figure~\ref{fig:expsetup} shows the basic setup to perform Synchrotron Radiation Interferometry (SRI). 
At ALBA, the SR is emitted by a dipole and the visible part is brought to the BL34 beamline using a set of 7 mirrors: one of which is in-vacuum and it is used to select just the visible light, the other mirrors are in atmospheric pressure. Once the light arrives at the beamline, the SR hits the NRA mask, and the interferometric pattern is imaged through a focusing lens and a magnifier onto the CCD. We have used 2, 3, 5 and 7-hole aperture masks~\cite{Carilli24}. 

\begin{figure}[!htb]
   \centering
   \includegraphics*[width=.95\columnwidth]{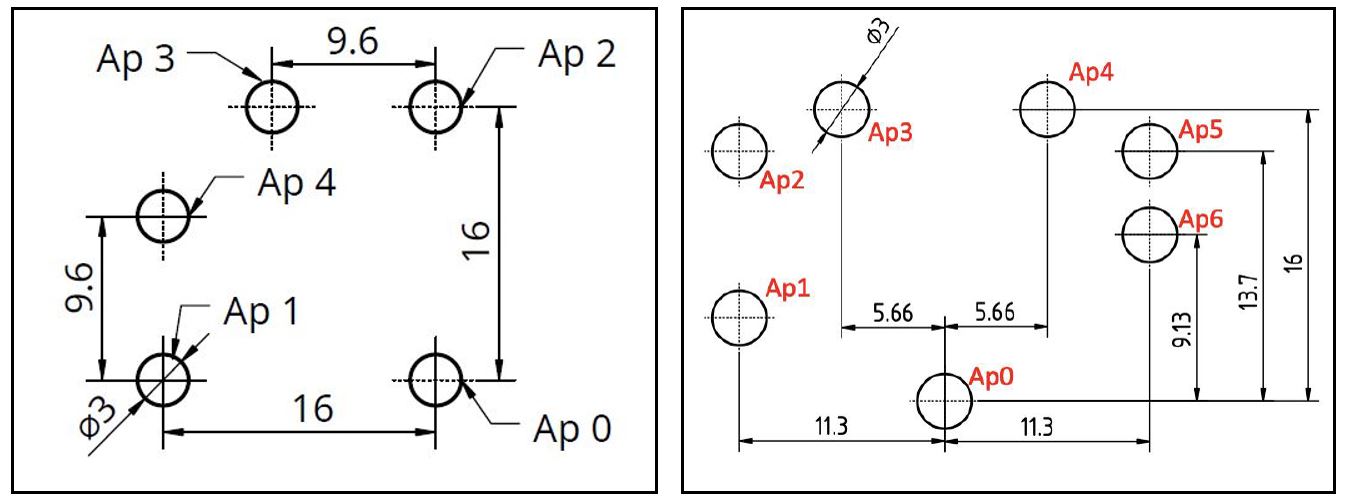}
   \caption{Sketch of the 5-hole NRA mask (left) and 7-hole mask (right). Distances are given in mm. }
   \label{fig:mask}
\end{figure}

Figure~\ref{fig:mask} shows the geometry of the 5-hole (left) and 7-hole mask (right), adapted from~\cite{Gonzalez:11}. All holes are 3~mm diameter. The 3-hole mask is basically the same as the 5-hole, but only uses Ap-0, 1 and 2. 

\section{Coherence Model}

\begin{figure}[!htb]
   \centering
   \includegraphics*[width=.95\columnwidth]{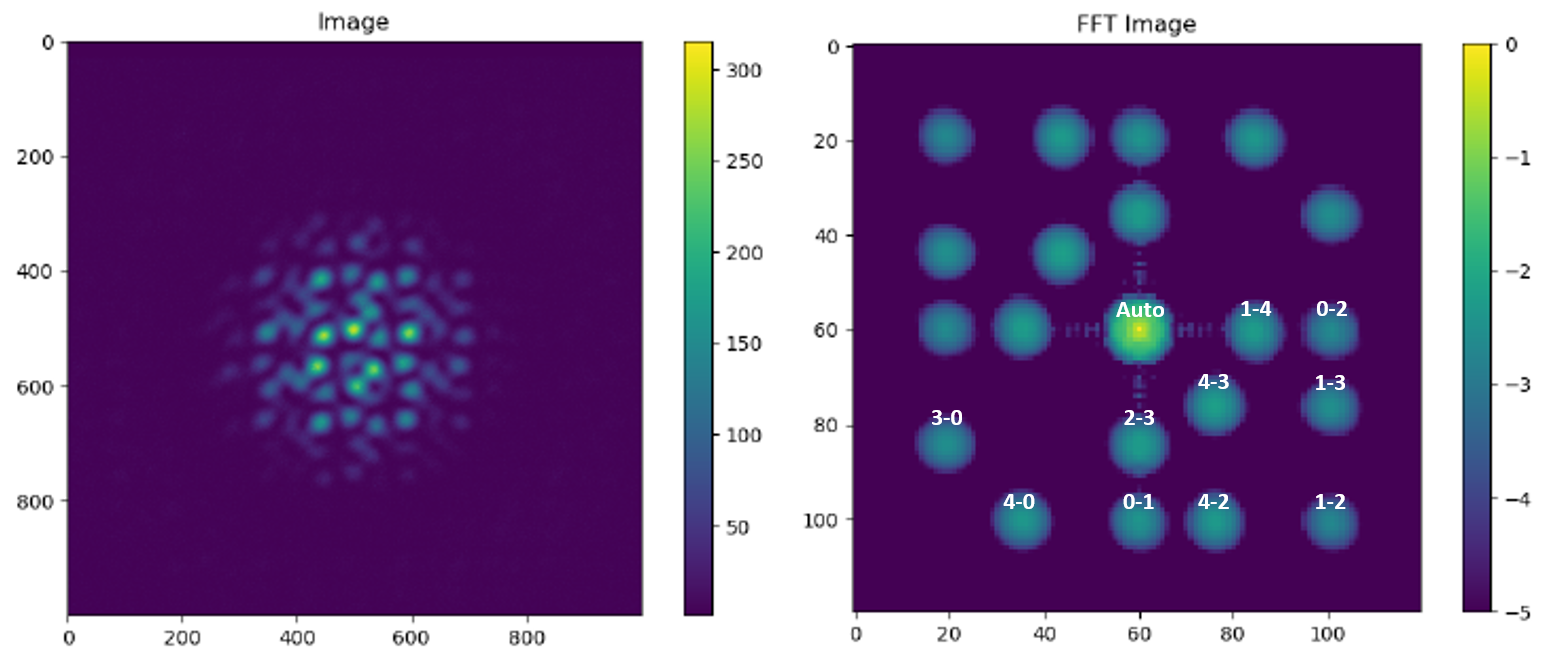}
   \caption{Raw interference image of the 5-hole NRA (left) and amplitude of the 2D FFT of it (right). Labels on the right plot identify peaks of amplitude with the pair of apertures in the mask from Fig.~\ref{fig:mask}. Color scale in the raw image is linear, while the 2D FFT plot is shown in log scale. }
   \label{fig:Imag:FFT}
\end{figure}

The interference pattern produced by a 5-hole NRA mask using a filter of $\lambda=$538~nm is shown in Fig.~\ref{fig:Imag:FFT}, left. 
As opposed to the 2-aperture interferometry, the interference pattern is quite complex and it is more convenient to work in the Fourier domain. The power spectrum (amplitude of the 2D FFT) is shown if Fig.~\ref{fig:Imag:FFT}, right. 
Since each baseline between every pair of apertures in the mask is unique, each spatial frequency in the Fourier domain is measured by only a single pair of apertures in the mask. 
Therefore, we can identify the discrete peaks in the Fourier space with the pair of apertures in the mask (see labels on Fig.~\ref{fig:Imag:FFT}, right).

The correlated power for each baseline (or the Visibility between apertures $i$ and $j$, $V_{\rm i,j}$) is calculated by the complex sum within the circle of the peak. The proper diameter (in pixels) is heuristically found, and it is typically 7 pixels for 3$\,$mm apertures mask.  
The Visibilites are a function of the degree of coherence $\gamma$ and the intensities $I$ going through each aperture $i$: 
\begin{equation}
\begin{split}
\|V_{ij}\| & =  \gamma_{\rm ij} \|G_i\| \|G_j\|~ , \\
  \|V_{\rm auto}\| & =  \sum_i  \|G_i\|^2 ~,
  \end{split}
  \label{eq:vis}
\end{equation}
\noindent
where $\gamma_{\rm ij}$ is the coherence at the Fourier coordinates  $u_{\rm ij}, v_{\rm ij}$ of the $i,j$ peak, $V_{\rm auto}$ is the central peak in Fig.~\ref{fig:Imag:FFT}, corresponding to the auto-correlated peak of each aperture. The amplitude of the complex voltage gains, $G_i$, are the square root of the intensity of the incident radiation power in the aperture $i$, so modulus $G_i = \sqrt{I_i}$. 

Following the van Cittert-Zernike (VCZ) theorem, the coherence function corresponds to the Fourier transform of the intensity distribution. If the electron beam follows a 2D Gaussian distribution, the coherence is also a 2D Gaussian with major axis, minor axis, and tilt angle described by:
\begin{align}
  \gamma(u,v) = V &\exp[- (a u^2 + b v^2 + 2cuv) ]~,
\end{align}
\noindent
where $V = V_{i,j} + V_{\rm auto}$ (see Eq.~\ref{eq:vis}), and $a,b,c$
are the parameters to be fitted and the 2D Gaussian parameters
$\sigma_u, \sigma_v$ and $\theta$ are calculated using the standard
equations in~\cite{sigma:wiki}. The final results for the source size
are inferred following the VCZ equation. In the horizontal case:
\begin{align}
  \sigma_x =  \frac{\lambda L}{2 \pi \sigma_u}~,
\end{align}
\noindent
where $\lambda$ is the light wavelength (in this case, (538$\pm$10~nm) and $L$ is the distane between the source point and the mask position. An analogous equation holds for the vertical case. 

\begin{figure}[!htb]
   \centering
   \includegraphics*[width=.7\columnwidth]{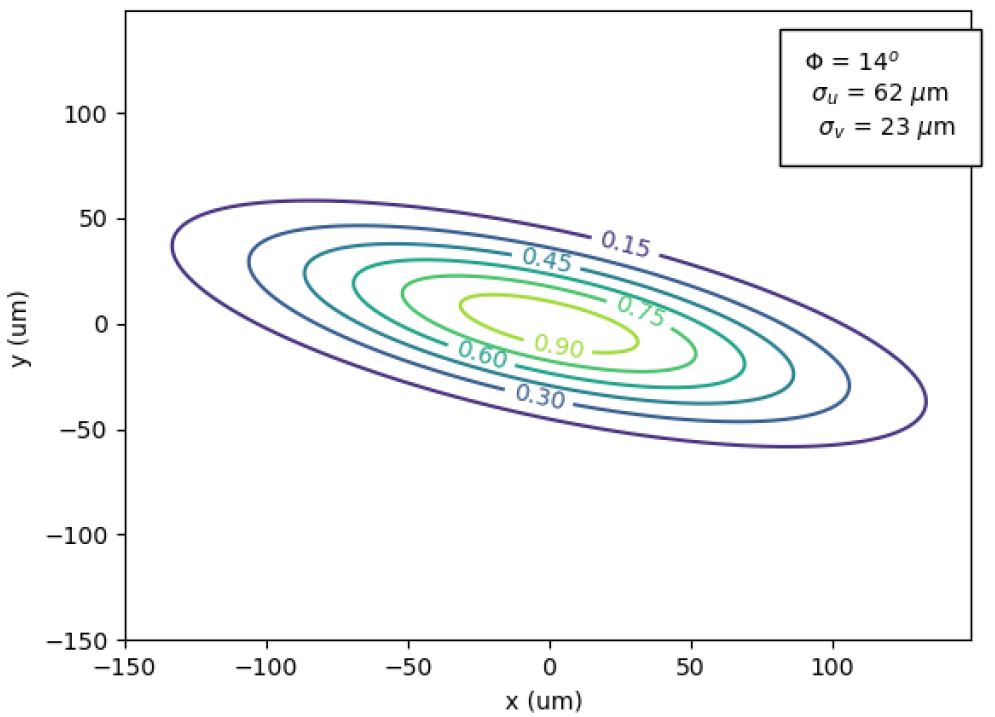}
   \caption{Beam ellipse obtained after the interferogram image in Fig.~\ref{fig:Imag:FFT}. }
   \label{fig:ellipse}
\end{figure}

Following this method, the maximum likelihood reconstructed 2D ellipse
corresponding to the interferogram in Fig.~\ref{fig:Imag:FFT} is shown
in Fig.~\ref{fig:ellipse}, which is in agreement with the results
obtained using the rotating 2-aperture mask and with the LOCO
analysis.

Note that this model includes 3 unknown parameters from the electron beam ($\sigma_x, \sigma_y, \theta$), plus the $N$ gains (or intensities) $G_i$ coming from the mask itself. 
For a N-hole mask, we have ($N(N-1)/2$) distinct peaks in the Fourier domain. Thus, with a 3-hole mask we have 6 unknown parameters but only 4 peaks in Fourier domain (3 cross correlations and 1 zero-spacing autocorrelation), therefore it is not enough to infer the whole beam with one interferogram. Instead, with a 5-hole non-redundant mask we have 8 unknown parameters but 11 peaks, and thus it is enough to fully reconstruct the 2D Gaussian beam.  
The fact that these gains are introduced in the fitting routine is what we call self-calibration and is the novelty of this method. 

We also stress that the mathematical analysis has some subtleties such as background subtraction, image padding, etc that are key ingredients to have optimum results. These details are thoroughly explained  in Refs.~\cite{Carilli24, BN:PRAB}. 




\section{Results}

\begin{figure}[!htb]
   \centering
   \includegraphics*[width=.95\columnwidth]{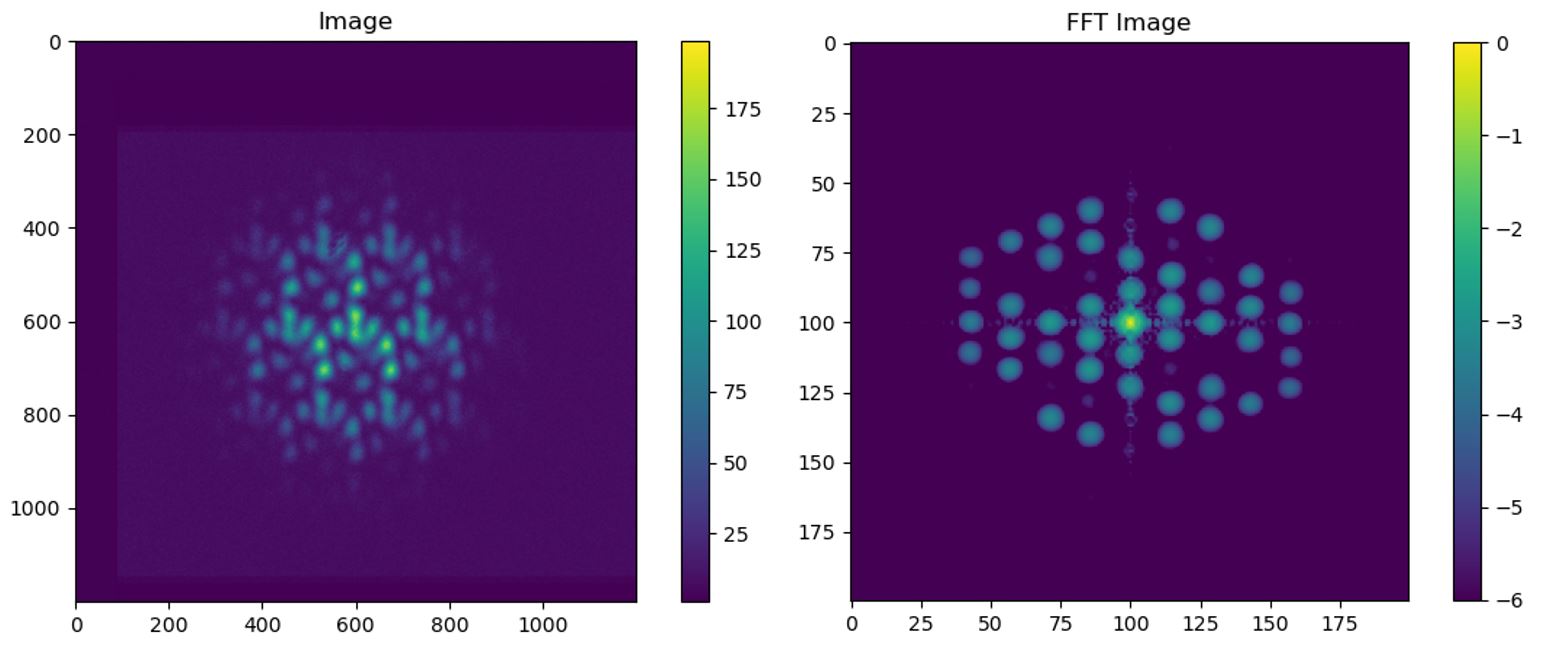}
   \caption{Raw image (left) of the 7-hole mask interferogram and amplitude of the 2D FFT of it (right). The mask geometry is shown in Fig.~\ref{fig:mask}, but the holes diameter has been reduced to 2~mm. Color scale in the raw image is linear, while the 2D FFT plot is shown in log scale.}
   \label{fig:7hole:img:ffg}
\end{figure}

In order to check the technique over a wide range of electron beam sizes, we have done a beam coupling scan using the skew quadrupoles in the ALBA storage ring. This allows us to change the vertical beam size from around \SI{12}{\um} to \SI{35}{\um}. 
The measurements are taken with a 7-hole mask, which introduces a larger coverage in the (u,v) plane, and therefore provide better constraints on the beam shape. Furthermore, we also use $\lambda$=400$\pm$10~nm to improve the resolution. Unfortunately, the peaks in the Fourier domain using the 7-hole mask (shown in Fig.~\ref{fig:mask}, right), sometimes present a small overlap. We solve this issue by reducing the mask hole diameters from 3 to 2~mm. An example of the interferogram and the power spectrum is shown in Fig.~\ref{fig:7hole:img:ffg}. 

\subsection{Exposure Time Scan}

In practice, the technical choice of parameters also become a key ingredient in the method. At ALBA, the visible light is brought from the source point to the experimental station using 7 mirrors. Except the first mirror, which is located in-vacuum and selects the visible part of the radiation, the rest of the mirrors are out-of-vacuum and supported in metallic bars. As a consequence, the light wavefront arriving to the mask suffers important vibrations~\cite{LT:IBIC14}. 

\begin{figure}[!htb]
   \centering
   \includegraphics*[width=.9\columnwidth]{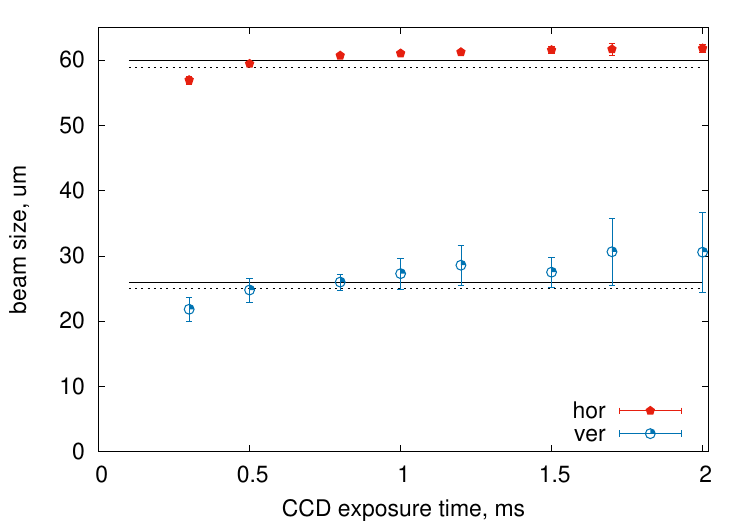}
   \caption{Exposure time scan using 7-hole mask with 2mm diameter. We mark the values inferred using the RSRI (dashed line) and using the emittance found in the ALBA pinholes (continuous black line). }
   \label{fig:ET}
\end{figure}

Thus it is important to chose a Exposure Time (ET) which provides enough SNR, but does not produce decoherence in the interference pattern. Figure~\ref{fig:ET} shows the horizontal and vertical beamsizes (actually, the major and minor axis of the beam ellipse) results after scanning the CCD exposure time. Each value corresponds to the average of 30 images, while the error bar corresponds to the rms of these 30 analysis. For comparison, we mark the values obtained using the 2-aperture rotating SRI (denoted as "RSRI" in the following) and the values calculated from the emittance calculated using the x-ray ALBA pinholes~\cite{UI:IBIC22}. 

The general tendency is that the beamsize increases with the exposure time. This indicates that there is decoherence produced by the aforementioned vibrations. 
Note also that in the vertical plane the spread in the results is quite big, and increases for exposure times larger than 1~ms, indicating that the vibrations may be on the order of 1~kHz. For these reasons, in the following we will do the coupling scan changing also the exposure time, but only with ET=0.3, 0.5 and 1~ms. 

\subsection{Coupling Scan}

The most challenging conditions are given in the vertical plane, where the beam is expected to be below \SI{15}{\um}. These produces a large coherence length, which is barely sampled using the mask in Fig.~\ref{fig:mask}. Unfortunately, the largest baseline is 16~mm, and cannot be larger because the wavefront arriving to the beamline is around 18~mm. This limitation arises from the in-vacuum mirror design, and cannot be easily changed. 
In these conditions, the visibilities are close to unity and the calculations become very sensitive to small errors, such as background or vibrations. 

\begin{figure}[!htb]
   \centering
   \includegraphics*[width=.9\columnwidth]{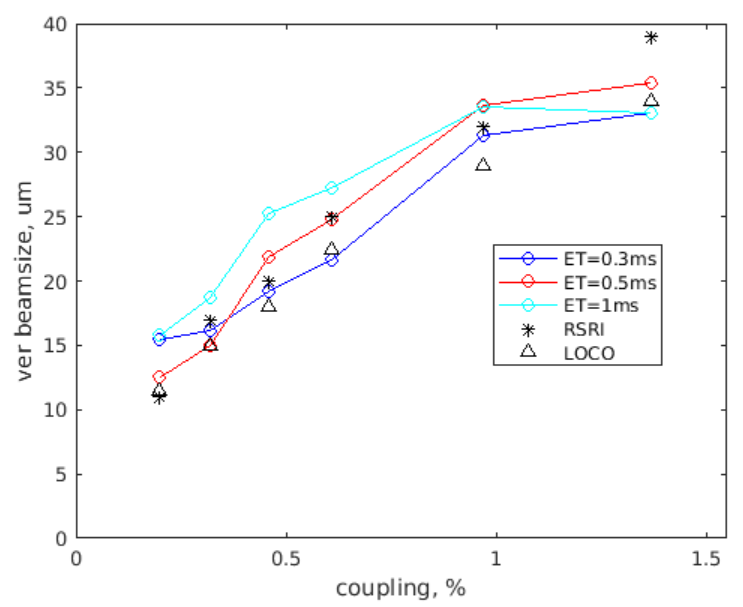}
   \caption{Vertical beamsizes measured with the 7-hole NRA for different exposure times, compared with the results given by the RSRI (black stars) and LOCO values (black triangles).}
   \label{fig:coupscan:ver}
\end{figure}

Figure~\ref{fig:coupscan:ver} shows the vertical measurements using the 7-hole NRA with ET=0.3, 0.5 and 1~ms, which are compared with the results obtained using the RSRI and LOCO. Note that the agreement is consistent for all ET, but a better agreement is shown for the ET=0.5~ms, especially for the smallest coupling which resulted in beamsizes smaller than \SI{15}{\um}. This is consistent with the results in the previous Exposure time scan (see Fig.~\ref{fig:ET}). 

\begin{figure}[!htb]
   \centering
   \includegraphics*[width=.99\columnwidth]{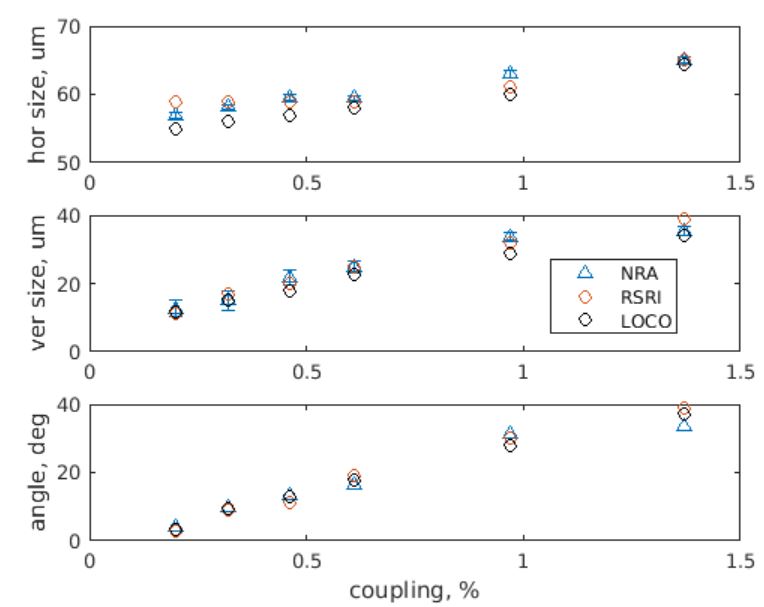}
   \caption{Comparison of the horizontal beamsize (top), vertical beamsize (middle) and tilt angle (bottom) of the electron beam while changing the beam coupling using the skew quadrupoles. We compare the data using the 7-hole NRA with values inferred from RSRI and LOCO.}
   \label{fig:coupscan:all}
\end{figure}

Finally, in Fig.~\ref{fig:coupscan:all} we show the results using the 7-hole NRA method, with ET=0.5\,ms, for all the parameters defining the 2D Gaussian: $\sigma_x$ (top), $\sigma_y$ (middle) and $\theta$ (bottom). The plot also compares the measurements with other results, obtained using the RSRI (red circles) and LOCO analysis (black circles). We note the good agreement in all cases. 

\section{CONCLUSION}

This reports presents a new method to fully reconstruct the 2D transverse distribution of the electron beam using the interferometry produced by the synchrotron radiation and a multi-hole Non-Redundant Aperture (NRA) mask. Using a minimum of 5-hole apertures mask, the 2D electron beam can be characterized with only one interferogram, which means the technique can be used for real-time applications (processing time less than a few milliseconds). In order to evaluate the technique for different beam sizes, the beam coupling in the machine has been varied using the skew quadrupoles. The measurements using the 7-hole NRA mask are in agreement with the results given by well-established methods (RSRI and LOCO). The 2D NRA SRI technique provided reliable measurements within a range of $\sim$\SI{10}{\um} (smallest achievable vertical beam size) to around \SI{70}{\um}.

\section{ACKNOWLEDGEMENTS}
The National Radio Astronomy Observatory is a facility of the National Science Foundation operated under cooperative agreement by Associated Universities, Inc.. Patent applied for: UK Patent Application Number 2406928.8, USA Applications No. 63/648,303 (RL 8127.306.USPR) and No. 63/648,284 (RL 8127.035.GBPR).

\end{document}